\documentclass[useAMS, usegraphicx, usenatbib]{mn2e}
\usepackage{times}  

\newcommand{\aap}   {A\&A}
\newcommand{\aj}    {AJ}
\newcommand{\apj}   {ApJ}
\newcommand{\apjl}  {ApJ}
\newcommand{\araa}  {ARA\&A}
\newcommand{\mnras} {MNRAS}
\newcommand{\pasj}  {PASJ}
\newcommand{\pasp}  {PASP}

\title[Collimation by a Disc Field]{Collimation of a Central Wind by a
Disc-Associated Magnetic Field}

\author[Matt, Winglee, \& B\"ohm]{Sean Matt$^{1}$\thanks{email:
matt@physics.mcmaster.ca; CITA National Fellow}, Robert
Winglee$^{2}$, and Karl-Heinz B\"ohm$^{3}$ 
\\ $^{1}$Physics \& Astronomy Department, McMaster University, Hamilton ON,
Canada L8S 4M1
\\ $^{2}$Earth \& Space Sciences, University of Washington, Seattle WA, 
U.S.A. 98195
\\ $^{3}$Astronomy Department, University of Washington, Seattle WA, 
U.S.A. 98195}

\begin{document}


\pagerange{\pageref{firstpage}--\pageref{lastpage}} \pubyear{2002}

\maketitle

\label{firstpage}

\begin{abstract}

Studies of jets from young stellar objects (YSO's) suggest that
material is launched from a small central region at wide opening
angles and collimated by an interaction with the surrounding
environment.  Using time-dependent, numerical magnetohydrodynamic
simulations, we follow the detailed launching of a central wind via
the coupling of a stellar dipole field to the inner edge of an
accretion disc.  Our method employs a series of nested computational
grids, which allows the simulations to follow the central wind out to
scales of tens of AU, where it may interact with its surroundings.
The coupling between the stellar magnetosphere and disc inner edge has
been known to produce an outflow containing both a highly collimated
jet plus a wide-angle flow.  The jet and wide-angle wind flow at
roughly the same speed (100--200 km s$^{-1}$), and most of the energy
and mass is carried off at relatively wide angles.  We show that the
addition of a weak disc-associated field ($\ll 0.1$ Gauss) has little
effect on the wind launching, but it collimates the entire flow (jet +
wide wind) at a distance of several AU.  The collimation is
inevitable, regardless of the relative polarity of the disc field and
stellar dipole, and the result is a more powerful and physically
broader collimated flow than from the star-disc interaction alone.
Within the collimation region, the morphology of the large-scale flow
resembles a pitchfork, in projection.  We compare these results with
observations of outflows from YSO's and discuss the possible origin of
the disc-associated field.

\end{abstract}


\section{Introduction} \label{sec_vertjet_intro}

Several decades of observing young stellar objects (YSO's) have
revealed a ubiquitous coincidence of structured outflows and accretion
discs, accompanying the star formation process.  See
\citet{reipurthbally01}, \citet{eisloffelea00}, and
\citet{koniglpudritz00} for detailed reviews of Herbig--Haro (HH)
objects and YSO outflows in general.  The outflows are invariably
collimated to some degree and always travel nearly perpendicular to
accretion discs.  Often, relatively wide-angle molecular (e.g., CO)
outflows escort narrower optical outflows from the same source.  The
optical outflows typically consist of jets, which are often resolved
into knots (HH objects), moving with bulk velocities of 100-200 km
s$^{-1}$ and occasionally exhibit high degrees of collimation.  The
molecular outflows are slower, moving a few tens of km s$^{-1}$.  The
question of whether CO outflows originate from physically distinct,
wide-angle winds \citep[e.g.,][]{kwantademaru88} or whether they
result from momentum transferred to molecular cloud material from the
jet \citep[e.g.,][]{koniglpudritz00} has not been answered
\citep{reipurthbally01, leryea02}.

\citet{kwantademaru88, kwantademaru95} review high spectral resolution
observations of the central regions of young stellar systems at the
origin of collimated outflows.  They point out that several
characteristics of optical forbidden emission lines are most easily
explained by the presence of two winds of different origin.  The two
winds consist of the following: 1) a fast ($\ga 100$ km s$^{-1}$) wind
originating from the very central region of the accretion disc or from
the star itself, and 2) a slow ($\sim$ 10 km s$^{-1}$) wind launched
from the outer ($> 10 R_\odot$) region of the accretion disc (the
`disc wind').  Evidence for at least two, physically distinct winds
exists for several objects \citep[e.g.,][]{solfbohm93,
fridlundliseau98, bacciottiea00, pyoea02}.

Prevalent theoretical models for winds launched from accretion discs
\citep[see, e.g.,][for a review]{shibatakudoh99,koniglpudritz00} hold
that the final wind velocity is of the order of the Keplerian
rotational velocity of the launch point.  The high velocity of
observed optical jets suggests that they originate from a deep
potential well \citep{kwantademaru88}, requiring the launching region
to be much less than an AU in extent, for reasonable parameters.  To
within observational limits, the jets often appear to emerge from
their source with relatively large widths (several 10's of AU), but
remain highly collimated immediately afterward
\citep[e.g.,][]{burrowsea96, eisloffelea00, reipurthea00,
reipurthea02}.  We therefore adopt the view that jets are launched
from a region within several stellar radii, initially with large
opening angles (hereafter the `central' or `wide-angle' wind), but
then become collimated within several 10's of AU.

In wind launching theory, the self-collimation of the wide-angle wind
(due to azimuthal magnetic fields carried in the wind itself) cannot
occur for all of the wind material \citep[][]{blandfordpayne82,
sakurai85, shuea95}.  In other words, without an external influence on
the wind, the launching models predict that there is always some
material leaving at very wide angles.  On the other hand, theoretical
studies of jets propagating into the surrounding environment most
often assume that the flow is completely collimated
\citep[][]{frank3ea02}.  Thus, there exists a gap in the theory
between the launching and the propagation of YSO outflows.  There is a
need to understand how the initially wide-angle flows interact with
their environment, presumably leading to a collimation of the entire
flow.

Recent work by \citet[][also see references therein]{gardiner3ea02}
considered the interaction of wide-angle winds with an unmagnetized,
collapsing environment.  They convincingly demonstrated that such
flows may be collimated by shock focusing, especially when azimuthal
magnetic fields are present in the wide-angle wind.  For the central
wind, they adopted a flow similar to that predicted by wind launching
theory on large scales, though they were unable to directly model the
launching of the flow.  As a result, all of the ambient material, even
that in the equatorial plane, was swept away in the wind, preventing
any further accretion into the central region.  This is in contrast to
the expectation from a system containing an accretion disc, which
prevents any outflow in a direction near the equatorial plane.  In
this paper, we consider the interaction of a wide-angle wind with a
magnetized environment.  Our method allows us to follow the details of
the launching of the wide-angle flow from a region very near the
central star, while still following the wind out to tens of AU, where
it interacts with its environment.

Since the central, wide-angle winds become collimated along the
rotational axis of the accretion disc, the collimation process itself
must be associated with the angular momentum of the system.  In this
paper, we use numerical magnetohydrodynamic (MHD) simulations to
explore the suggestion of \citet{kwantademaru88} that disc-associated
magnetic fields collimate a central, fast wind into an optical jet.
Large-scale disc fields may be generated by currents in the disc
\citep[][]{spruit3ea97}, or they could be embedded in the surface of
the disc and carried outward in a slow disc wind
\citep[e.g.,][]{blandfordpayne82, uchidashibata85, pudritznorman83,
pudritznorman86, shibatauchida86, kwantademaru88, shibatauchida90,
ouyedpudritz97, ustyugovaea99}.  We adopt an initially vertical field
to approximate the magnetic configuration expected from a disc on
scales that are large compared to the wind launching region, but small
compared to the size of the disc \citep[e.g.,][]{blandfordpayne82,
spruit3ea97}.  The disc fields we consider are very weak compared to
the ones that are likely to be involved in launching the central
outflow (e.g., fields near the surface of the star).  We will show
that for typical YSO outflow parameters, disc fields of much less than
0.1 Gauss are capable of completely collimating the flows.

Our approach is to use the 2.5D (axisymmetric), resistive MHD code
described by \citet{mattea02} to examine, parametrically, the effect
of a vertical field on a wide-angle outflow.  In order to gain insight
into the physics involved, we first examine (in \S
\ref{sec_vertjet_isotropic}) the simplified situation of an isotropic,
unmagnetized central wind interacting with a stagnant environment that
contains a constant magnetic field.  The main result is that
collimation is inevitable, and that the size scale of collimation is
inversely proportional to the strength of the field.  In section
\ref{sec_vertjet_application}, we assess the effect of a vertical
field on a more realistic (anisotropic) outflow launched from the
central region of a YSO.  Our simulations are able to resolve the
self-consistent interaction between the stellar dipole magnetic field
and the inner edge of an accretion disc, which results in a highly
structured, wide-angle flow from the central region.  This central
wind launching mechanism has been studied by several authors and is
reviewed in section \ref{sub_vertjet_emireview}.  Using a system of
nested computational grids, we follow this central flow to large
distances, where it interacts with the vertical disc field.  The
interaction of this complex central wind with the disc field results
in a highly structured flow, in contrast to the case with an isotropic
central wind.  We compare our results with observations, discuss the
possible origin of a disc field, and suggest possible improvements to
the model in section \ref{sec_vertjet_discussion}.

\section{Collimation of an Isotropic Wind} \label{sec_vertjet_isotropic}

In order to get a preliminary idea of the physics controlling the
collimation of a central wind by a constant, vertical field, we first
examine a simplified situation.  Consider an isotropic wind that
collides with a stagnant, constant density environment.  In the
absence of any magnetic fields, the result will be the well-studied
case of a spherical wind-blown bubble \citep[WBB;][]{kwok3ea78,
dysonwilliams80}.  In that case, the wind acts as a piston pushing
against the ambient medium, and an expanding structure results with
two shocks and four distinct zones (free-flowing wind, shocked wind,
shocked ambient, and undisturbed ambient material).

The ambient material acts as an inertial barrier to the wind.  When
ordered magnetic fields are present, they introduce a directional
dependence to the inertial barrier.  Consider, in particular, the
simple case of a constant magnetic field permeating all of space and
directed along the cylindrical $z$-axis ($B_z$ = constant).  The wind
flowing along the axis will experience no magnetic force in the $z$
direction.  However, the wind flowing perpendicular to the field
(i.e., in the equatorial plane, $z$ = 0) will be impeded by the
magnetic pressure (assuming the wind is sufficiently ionized).  If the
field is relatively weak and unimportant near the source of the wind,
it will be swept outward, in front of the wind.  The dynamic pressure
of the wind decreases with radius, so there will always exist a
location at which the constant field becomes dynamically important and
completely stops the wind.  In the case of a wind that is constant in
time, continuity requires that the wind stagnating at the equator be
deflected toward the $z$-axis.

This relatively simple system, consisting of an isotropic central wind
interacting with a plane-parallel magnetic field, has been studied by
a several authors \citep{levy71, konigl82, ferriere3ea91,
stonenorman92}.  They were able to predict the shape and expansion of
the bubble of swept up ambient material, though they did not follow
the details of the flow interior to that WBB.  In this paper, we are
interested in the collimation of the material originating in the wind,
so this section contains a relatively simple study of this system and
focuses on the conditions in the flow interior to the WBB.

For simplicity, assume that the wind has a constant velocity that is
much higher than its own sound and magnetic wave (Alfv\'en) speeds.
Such is the case in most astrophysical outflows beyond several times
the size of the launching region.  Also, assume the external medium
has a constant temperature and vertical magnetic field, $B_z$, such
that the Alfv\'en speed is much higher than the local sound speed.
Mathematically,
\begin{eqnarray}
v_{\rm w} \gg c_{\rm s,w}, ~~~~~
v_w \gg v_{\rm A,w}, ~~~~~
v_{\rm A,a} \gg c_{\rm s,a}, 
\label{eqn_vertjet_assump}
\end{eqnarray}
where $v_{\rm w}$ is the wind speed, $c_{\rm s}$ is the sound speed,
$v_{\rm A}$ is the Alfv\'en speed, and the subscripts `w' and `a'
refer to the value in the wind and ambient material, respectively.
The conditions (\ref{eqn_vertjet_assump}) mean that the kinetic energy
dominates in the free-flowing wind (thermal and magnetic energy are
unimportant), and the magnetic energy dominates in the ambient medium
(so that thermal pressure is unimportant; also neglecting the effects
of gravity, which is valid far from the wind acceleration region).

Consider an initially stagnant medium and a spherical wind that turns
on at $t$ = 0.  The wind will push against a weak magnetic field,
`carving out' a field-free region.  The magnetic field will become
important when the ram pressure in the wind ($\rho v_{\rm w}^2$, which
falls of with distance from the star) is equal to the magnetic
pressure of the external environment.  For an isotropic wind, the
balance of these two pressures on the equator (cylindrical $z = 0$) is
equivalent to the balance of the ambient magnetic and wind kinetic
energy densities.  Therefore, the key parameter for determining the
importance of the magnetic barrier ($B_z$) to the wind is simply the
ratio
\begin{equation}
\sigma = {{B_z^2 ~ R^2} \over {\dot M_{\rm w} ~ v_{\rm w}}},
\label{eqn_vertjet_sigma}
\end{equation}
where $\dot M_{\rm w} = 4 \pi R^2 \rho v_{\rm w}$ is the mass outflow
rate in the wind.  Note that $\sigma$ increases as the square of
spherical radius ($R$) so that, even if $B_z$ is very small, magnetic
forces will always be important at some distance.

When the expanding shell of shocked material reaches a radius at which
$\sigma$ is near unity, the shell will deviate from spherical
symmetry.  The cylindrical $r$ component of the wind velocity will be
forced to zero.  Since the wind travels faster than any
information-carrying waves, the deceleration of the wind will be in
the form of a shock.  Pressure forces behind the shock will channel
the post-shock flow toward the axis.  The location of the deceleration
shock in the equatorial plane is determined by setting $\sigma = 1$ in
Equation \ref{eqn_vertjet_sigma} and solving for $R$.  Our simulations
(presented in \S \ref{sub_vertjet_method1} and \S
\ref{sub_vertjet_results1}) agree with this prediction for various
wind and ambient parameters subject to conditions
\ref{eqn_vertjet_assump}.

	\subsection{MHD Simulations} \label{sub_vertjet_method1}

To test the simple concept of an isotropic central wind interacting
with a constant vertical field, and to develop insight into more
complex problems, we have carried out a number of numerical
simulations.  We use the 2.5D MHD code of \citet{mattea02}, and the
reader will find the details there \citep[see also][]{goodson3ea97,
matt02}.  The simulations employ a two-step Lax-Wendroff
(finite-difference) scheme to solve the ideal MHD equations on a
series of nested boxes.  The equations allow for an ideal gas equation
of state ($\gamma$ = 5/3) and assume that the system is axisymmetric
($\partial/\partial \phi = 0$ for all quantities).  Each box consists
of a $101 \times 100$ cylindrical ($r$-$z$) grid with constant grid
spacing.  The boxes are nested concentrically, so that the inner box
represents the smallest domain at the highest resolution.  The next
outer box represents a twice the domain size with half the spatial
resolution (and so on for other boxes).  For the simplified approach
of this section, we use five boxes and ignore the effects of gravity
and magnetic diffusion.  Outflow boundary conditions (all variables
are continuous) exist on the right and top boundary of the outermost
box, allowing material to leave the box without reflecting information
inward.

The wind source is a multiple layered boundary, consisting of a circle
with an outer radius of 10.5 grid points in the smallest box, centered
on the origin of the grid.  The boundary conditions on the wind sphere
are as follows: All quantities are fixed (velocity = 0.0, {\boldmath
$B$} = 0.0, and density and pressure are constant) for $R \le 8.5$ ($R
= (r^2+z^2)^{1/2}$).  For $8.5 < R \le 9.5$, the same is true except
that {\boldmath $B$} floats (i.e., it evolves naturally in the course
of the simulations).  For $9.5 < R \le 10.5$, the velocity is fixed at
a constant, nonzero value and is directed radially, while all other
quantities are the same as for $8.5 < R \le 9.5$.  The existence of a
layer with a zero velocity and a floating {\boldmath $B$} interior to
a layer with nonzero velocity is necessary to preserve $\nabla \cdot
\mbox{\boldmath $B$} = 0$ near the wind source boundary (similar
conditions are used by \citealp{mattea00}).  The wind source in the
simulations represents a reference surface far from the region where
the wind is accelerated.  In this way, the fact that the wind is
supersonic and super-Alfv\'enic and the neglect of gravity is
justified.

The grid initially has a constant density, pressure, $B_z$, and zero
velocity.  The key parameter is $\sigma_*$, where the asterisk
subscript refers to the value at the radius of our wind source,
defined by $\sigma_* = B_z^2 R_*^2 \dot M_{\rm w}^{-1} v_{\rm
w}^{-1}$.  At $t$ = 0, the wind begins to move outward from the wind
source, pushing the ambient material ahead of it and sweeping up the
ambient magnetic field.  We first verified that, when $B_z$ = 0
($\sigma_* = 0$), the resulting WBB is spherically symmetric.  Nine
other cases, with finite $B_z$, were sufficient to cover a range of
parameters.  Table \ref{tab_vertjet_sim1} contains the parameter
values of each case, and the simulation results are the topic of the
next section.  We consider the case with $\sigma_*$ = 0.0087, wind
source to ambient density ratio ($\rho_*/\rho_{\rm a}$) = 10, wind
Mach number ($v_{\rm w}/c_{\rm s,w}$) = 5, and ambient Alfv\'en to
sound speed ratio ($v_{\rm A,a}/c_{\rm s,a}$) = 5, as our baseline,
with which to compare all other simulations.

	\subsection{Results} \label{sub_vertjet_results1}

In the very earliest stage of the simulations, the interaction between
the wind and the stagnant environment produces an expanding,
spherically symmetric WBB.  The flow has the following important
features: 1) an interior filled with free-flowing wind, 2) a reverse
shock (RS), where wind kinetic energy is converted into thermal
energy, 3) a contact discontinuity (CD) separating the shock-heated
wind gas from the shock-heated ambient gas, and 4) a forward shock
where the expanding structure runs into ambient material and heats it.
As the simulation evolves, and the structure expands, the vertical
magnetic field ($B_z$) in the ambient material is swept aside in $r$.
When the WBB reaches a location where $\sigma$ (Eqn.\
\ref{eqn_vertjet_sigma}) is near unity, the RS slows and eventually
stagnates.  The position of the CD on the equatorial plane also stops,
and pressure gradients in the region between the RS and CD channel
material toward the axis.  The end result is a steady-state,
collimated flow (a jet) propagating along the $z$-axis.

\begin{figure}
\centerline{\includegraphics[width=20pc]{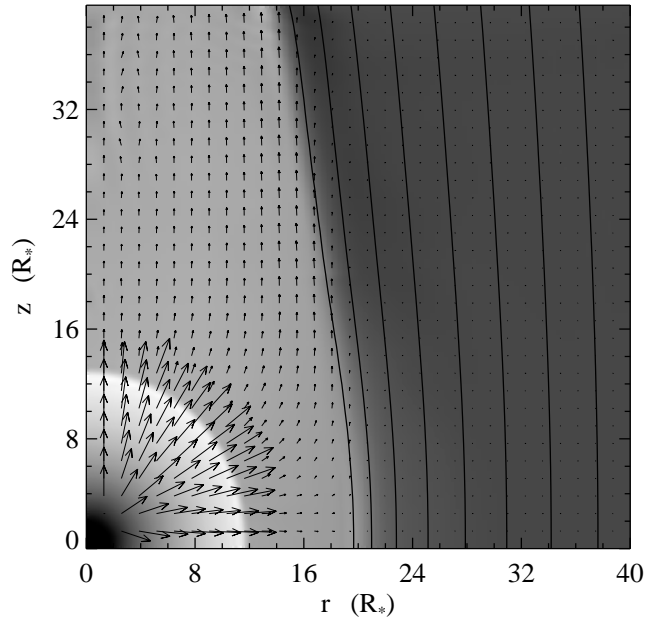}}

\caption[Density, Field Lines, and Velocity Vectors for the Baseline 
Case]{Greyscale image of $\log$ density (darker is higher density),
field lines, and velocity vectors illustrate the near steady-state
flow for the $\sigma_* = 0.0087$, baseline case. \label{fig_sigma1rho}}

\end{figure}

Figure \ref{fig_sigma1rho} illustrates the flow for the baseline case
near steady-state, within 40 wind source radii.  The figure clearly
demonstrates the existence of a RS at about 12 $R_*$ (where the
velocity of the wind suddenly drops) and the channeling via pressure
gradients of the post-shock flow toward the axis.  The deflected and
compressed field lines (near the CD) provide a $\mbox{\boldmath $J$}
\times \mbox{\boldmath $B$}$ force that balances the dynamic pressure
of the wind.  Figure \ref{fig_sigma1rho} confirms the conceptual model
of \citet[][compare with their fig.\ 2]{kwantademaru88}.

\begin{figure}
\centerline{\includegraphics[width=20pc]{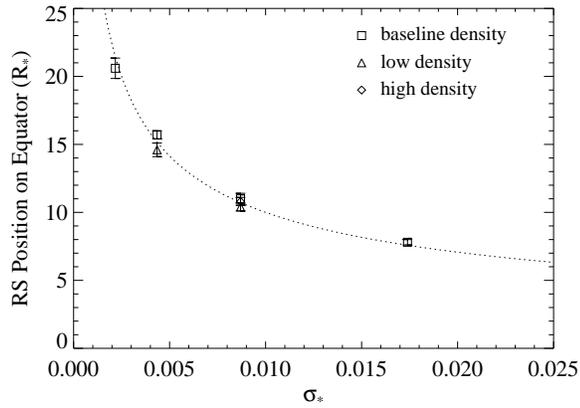}}

\caption[Location of Standing Shock for Various Cases]{Equatorial 
location of the standing, star-facing shock as a function of
$\sigma_*$.  The simulated data (symbols for various cases indicated
on figure) agree with the analytical prediction (dotted line).  The
errors in the simulated values (vertical bars) are due to the finite
width of the standing shock (due to numerical resolution).
\label{fig_rsvsigma}}

\end{figure}

We plot the equatorial stagnation location of the RS in seven of our
simulations, as a function of the parameter $\sigma_*$, in figure
\ref{fig_rsvsigma}.  Square symbols represent the four cases with
baseline ambient density (top four rows in table
\ref{tab_vertjet_sim1}), though each include a unique value of $B_z$.
The triangles represent two cases with the ambient density decreased
by a factor of 10 (row 5 and 7 of table \ref{tab_vertjet_sim1}), and
the diamond is for the case with increased ambient density (row 6).
We did not plot the results for the two simulations with various wind
and ambient temperatures (rows 8 and 9), since they were
indistinguishable from the baseline case.  The error bars in figure
\ref{fig_rsvsigma} represent the uncertainty due to the finite width
of the shock, caused mainly by the relatively low numerical
resolution.  The dotted line represents the radial location where
$\sigma$ = 1.0 for each $\sigma_*$.  Figure \ref{fig_rsvsigma}
demonstrates that, within the uncertainty, the location of the RS on
the equator is indeed predicted by equation \ref{eqn_vertjet_sigma}.

\begin{figure}
\centerline{\includegraphics[width=20pc]{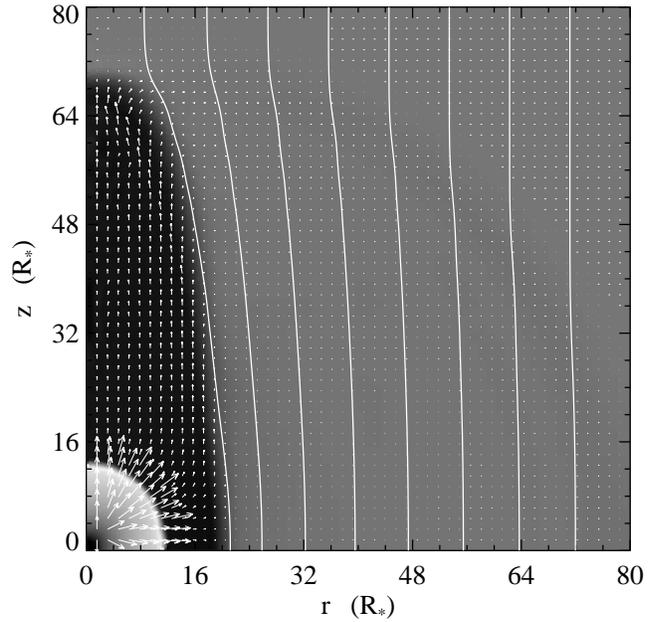}}

\caption[Pressure, Field Lines, and Velocity Vectors for the Baseline 
Case]{Greyscale image of $\log$ pressure (darker is higher pressure),
field lines, and velocity vectors illustrate the flow at a late time
and a larger scale for the $\sigma_* = 0.0087$, baseline case shown in
figure \ref{fig_sigma1rho}.  \label{fig_sigma1pre}}

\end{figure}

Figure \ref{fig_sigma1pre} is a greyscale image of the pressure
(logarithmic) for the baseline case.  The data are from the same time
as in figure \ref{fig_sigma1rho}, but they represent the next outer
box.  The figure demonstrates the propagation of the jet along the
axis.  The jet is composed of shocked gas, and it is supported
laterally by the balance of internal thermal pressure versus external
magnetic forces.  The outer edge of this jet (i.e., the edge of the
high pressure region) represents the CD and is thus the location of
the expanding bubble studied by (e.g.) \citet{stonenorman92}.  Though
the jet propagates at a velocity that is higher than $c_{\rm s,a}$ and
comparable to $v_{\rm A,a}$, it is traveling slower than its own
internal sound speed.  In the more general case of an anisotropic
central wind, it is possible for the shocked gas (the jet) to travel
faster than its own internal sound speed.  Also, if the gas were
allowed to cool radiatively, the jet structure would be different.

Since the flow is subsonic between the RS and CD, the pressure is
roughly constant and equal to the decrease in the wind dynamic
pressure across the RS (i.e., the kinetic energy in front of the shock
is converted to thermal energy behind it).  This is evident in figure
\ref{fig_sigma1pre}.  The thermal pressure of the shocked wind extends
the influence of the wind dynamic pressure to larger radii.  The width
of this extension on the equator (equal to the distance between the RS
and the CD) depends upon how fast the shocked wind is moved toward the
axis by small pressure gradients, which depends on how fast the
material on the axis flows away (i.e., the speed of the jet).

\begin{figure}
\centerline{\includegraphics[width=20pc]{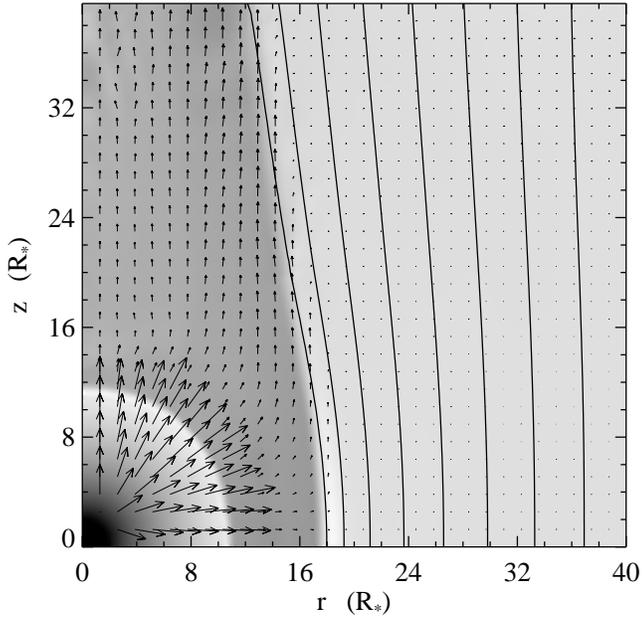}}

\caption[Density, Field Lines, and Velocity Vectors for the Low Density 
Case]{Greyscale image of $\log$ density (same greyscale as fig.\
\ref{fig_sigma1rho}), field lines, and velocity vectors illustrate the
flow at near steady-state for the $\sigma_* = 0.0087$ case with ambient
density a factor of 10 less than the baseline case (fig.\
\ref{fig_sigma1rho} and \ref{fig_sigma1pre}).
\label{fig_sigma1lowrho}}

\end{figure}

For given wind parameters and $B_z$, the jet speed is determined by
the conditions \citep[in particular, the Alfv\'en speed;
see][]{levy71} in the ambient medium.  The ambient density,
$\rho_{\rm a}$, is important, since it behaves as an inertial barrier
to the jet and it determines the Alfv\'en speed.  For a smaller
$\rho_{\rm a}$, the jet moves faster, allowing the equatorial region
between the RS and CD to be `drained' more quickly by pressure
gradients.  Figure \ref{fig_sigma1lowrho} shows the case that is
identical to the baseline case but with an ambient density decreased
by a factor of 10 (row 5 of table \ref{tab_vertjet_sim1}).  The figure
depicts an earlier time than figure \ref{fig_sigma1rho} because the
jet moves more rapidly toward the top of the outer box (which limits
the length of time the simulations can run), but a steady-state
already exists within about 20 wind sphere radii.  A comparison
between figures \ref{fig_sigma1rho} and \ref{fig_sigma1lowrho} reveals
a dependence of the CD location (equivalent to where the magnetic
field jumps in value) on the ambient density.

\begin{figure}
\centerline{\includegraphics[width=20pc]{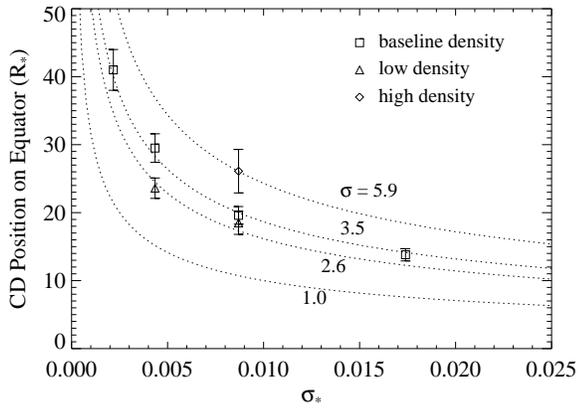}}

\caption[Location of the $B_z$ Jump for Various Cases]{Equatorial
location of the jump in $B_z$ as a function of $\sigma_*$.  The
simulated data (symbols for various cases indicated on figure) follow
the predicted trend for the standing shock (dotted lines for different
values of $\sigma$, chosen to fit the data) but also reveal a
dependence on conditions of the ambient material.  The errors in the
simulated values (vertical bars) are due to the finite width of the
jump in $B_z$ (due mainly to numerical resolution).
\label{fig_rbvsigma}}

\end{figure}

The relationship between the CD location and $\rho_{\rm a}$ is even
more apparent in figure \ref{fig_rbvsigma}, a plot of the CD location
versus the parameter $\sigma_*$.  The symbols represent the same cases
as in figure \ref{fig_rsvsigma}, and the dotted lines represent the
radii of various values of $\sigma$ as a function of $\sigma_*$.  For
the cases of any given ambient density, the points follow the same
trend (though shifted in position) as for the location of the RS.  For
an increased ambient density, the steady-state location of the CD on
the equator moves outward.

\section{Collimation of a YSO Outflow} \label{sec_vertjet_application}

In this section we consider the more complex case of an anisotropic
wind collimated by an initially vertical magnetic field embedded in an
accretion disc.  It is our goal to compare these results to
observations of outflows from YSO's, so we adopt a central wind
launching mechanism that is appropriate for those systems.  In the
absence of any disc field, this MHD mechanism (reviewed in \S
\ref{sub_vertjet_emireview}) produces an outflow with a collimated
(jet) component plus an uncollimated wind.  Both components of the
flow are relatively fast (100--200 km s$^{-1}$) and of the order of
the Keplerian rotation speed at the central region from which the wind
is launched.  Our simulations (described in \S
\ref{sub_vertjet_method2} and results presented in \S
\ref{sub_vertjet_results2}) resolve the detailed launching of the
central wind, and they will show that a large-scale field, embedded in
the accretion disc, will collimate the wide-angle component of the
wind.  This results in a more powerful and physically broader jet than
produced in the absence of the disc field.  We also show that the
simple results of the last section hold, even for the more complex
outflow morphologies produced in this section.

For simplicity, we will assume an initially constant, vertical field
frozen-into the disc and permeating an environment above and below the
disc that is initially in hydrostatic equilibrium.  The small
time-scales involved in properly capturing the launching of the
central wind make it impractical for us to also follow the launching
of disc plasma along the initially vertical field lines.  Therefore,
the existence of a wide-angle, slow, disc wind observed in some
systems \citep{kwantademaru88} is not addressed by our models.  Such a
wind may result from, or be responsible for, the existence of a
large-scale, primarily vertical magnetic field \citep[assuming it is
sufficiently ionized; e.g.,][]{blandfordpayne82}.  Our simple
treatment of the disc field will, in some ways, emulate a magnetized
disc wind that can act as a channel for the interior, wide-angle wind
\citep[e.g.,][]{ouyedpudritz97}.  We propose that the relatively broad
jet, produced by the collimation of a central wind by this disc field,
can explain some of the observational properties (e.g., the mass flux,
velocity, energy and momentum budget, and collimation) of optical
jets.


	\subsection{The Central Wind} \label{sub_vertjet_emireview}

For the central wind, we adopt a model studied by
\citet{hayashi3ea96}, \citet{goodson3ea97}, and \citet{millerstone97},
in which a rotating star couples to the inner edge of a conductive
accretion disc via a stellar dipole magnetic field.  In this model,
differential rotation between the star and disc twists up the magnetic
field, adding a $B_\phi$ component.  The magnetic pressure associated
with $B_\phi$ causes the stellar magnetosphere to expand (or
`inflate') above and below the disc.  Plasma loaded onto the
inflating field lines is driven outward in a burst.  When the
inflation speed becomes larger than any plasma wave speeds, the field
lines become effectively open, and material is magnetocentrifugally
launched from the disc inner edge via the Blandford--Payne mechanism
\citep{blandfordpayne82}.  Plasma launched from the disc inner edge
carries away angular momentum, so the disc spins down and moves
inward.  The inward motion of the disc forces together the open field
lines from the star and disc of opposite polarity, instigating a
reconnection of those lines.  After reconnection, material at the disc
inner edge is again magnetically coupled to the star, but with less
angular momentum than before, so that it accretes via funnel flow
along field lines onto the star.  The accretion of material unloads
the stellar magnetosphere, allowing it to expand outward and diffuse
into the new inner edge of the disc.  When the field begins to thread
the disc inner edge, differential rotation again twists up the field,
and the process repeats. This process of episodic magnetospheric
inflation (EMI) regulates both the accretion and ejection of material
within the region of the disc inner edge.


By following the outflow to several AU, \citet{goodson3ea97,
goodson3ea99} showed that the EMI mechanism produces a jet (hereafter,
the `EMI jet') that has knots associated with each reconnection event.
Note that the typical oscillation period for the EMI mechanism is
about 20--30 days, too short to explain the knot spacing in observed
jets \citep[e.g., in HH 30][]{burrowsea96}.  The rest of the flow
travels out at a wide opening angle, shocking the surface of a flared
accretion disc.  The flow speed for the EMI jet and wide-angle wind
component is $\ga$ 100 km s$^{-1}$, in rough agreement with
observations \citep{reipurthbally01}.  The energetics and the mass
outflow rate in the EMI wind are also within the range observed in
jets, but most of the kinetic energy and mass in the EMI wind is
contained in the wide-angle component (not in the jet).  In addition,
the width of the EMI jet component is a few AU, an order of magnitude
smaller than observed \citep[e.g.,][]{burrowsea96}.

In their limited parameter study of the EMI mechanism,
\citet{mattea02} showed that a weak magnetic field, aligned with the
rotation axis and threading the accretion disc, did not affect the
launching of the EMI outflow.  This should be true, as long as the
stellar dipole field is stronger than the vertical field at the inner
edge of the disc.  Their work focused on the integrated outflow
properties, rather than the details of the wind morphology, and they
were unable to address the collimation of the outflow by the vertical
fields because the size of their largest simulation grid reached only
to 0.75 AU.  In this section, we follow up the work of
\citet{mattea02} and, using insight gained from section
\ref{sec_vertjet_isotropic}, we will show that when one follows the
outflow to large enough distances, even a weak field will collimate
the entire EMI outflow.

Many other authors \citep[e.g.,][]{shuea94, lovelace3ea95,
fendtelstner99, kuker3ea03} have discussed outflows produced by the
interaction between a stellar magnetosphere and the inner edge of the
disc.  The EMI mechanism operates only when the magnetic diffusivity
is relatively small in these systems \citep{goodsonwinglee99}, but the
outflows produced in all of these studies contains a component that
flows out at wide angles.  We choose to work with the EMI model as one
example, but our conclusions apply to any wide-angle, central flow.

	\subsection{Simulation Method} \label{sub_vertjet_method2}

As in section \ref{sec_vertjet_isotropic}, we use the 2.5D MHD code of
\citet{mattea02}.  For the more complex simulations carried out in
this section, the code solves the ideal MHD equations with the added
physics (via source terms in the momentum, energy, and induction
equations) of gravity and Ohmic diffusion.  We also adopt their
baseline simulation case, in which a 1 $M_\odot$, 1.5 $R_\odot$
T-Tauri star with a 1.8 day rotation period interacts with a flared
accretion disc \citep[after the $\alpha$-disc model
of][]{shakurasunyaev73} via an axis aligned dipole magnetic field (=
909 Gauss on equator).  The reader will find more details of the
simulations and initial conditions in \citet{mattea02} and
\citet{goodson3ea97, goodson3ea99}.  Our simulations differ from those
of \citet{mattea02} only in that we use ten nested boxes, while they
used only four.  Each simulation box consists of a $401 \times 100$
cylindrical ($r$-$z$) grid, and the boxes are nested concentrically
(see \S \ref{sub_vertjet_method1}).  In this way, we fully capture the
launching of the outflow via the EMI mechanism in the innermost box,
while our outermost box follows the flow out to 24 AU.

First, we simulated the baseline case and follow the outflow to large
radii, allowable by the increased number of boxes.  We then ran four
other cases with an additional vertical and constant magnetic field
($B_z$) initialized everywhere in the simulation region.  All of the
cases are identical except for the strength of the initial vertical
field.  So the baseline case has $B_z$ = 0, and the others have
$|B_z|$ = 0.050, or 0.025, Gauss.  The vertical field can interact
(e.g., reconnect) with the field carried in the outflow, so there is a
need to assess the effect of polarity of the vertical field.  Thus,
for each case with nonzero $|B_z|$, we ran two cases, each with
opposite polarity.  Table \ref{tab_vertjet_sim2} lists the values of
$B_z$, the simulation time at which the simulations were stopped
($t_{\rm f}$), and some notes for each case.

Our choice of vertical field strengths was based on the need for the
field to be strong enough to collimate the flow within the outer box
and weak enough not to effect the EMI mechanism (which gives an upper
limit of $\sim 0.1$ G).  We used equation \ref{eqn_vertjet_sigma} to
estimate the required field.

	\subsection{Results} \label{sub_vertjet_results2}

\begin{figure*}
\centerline{\includegraphics{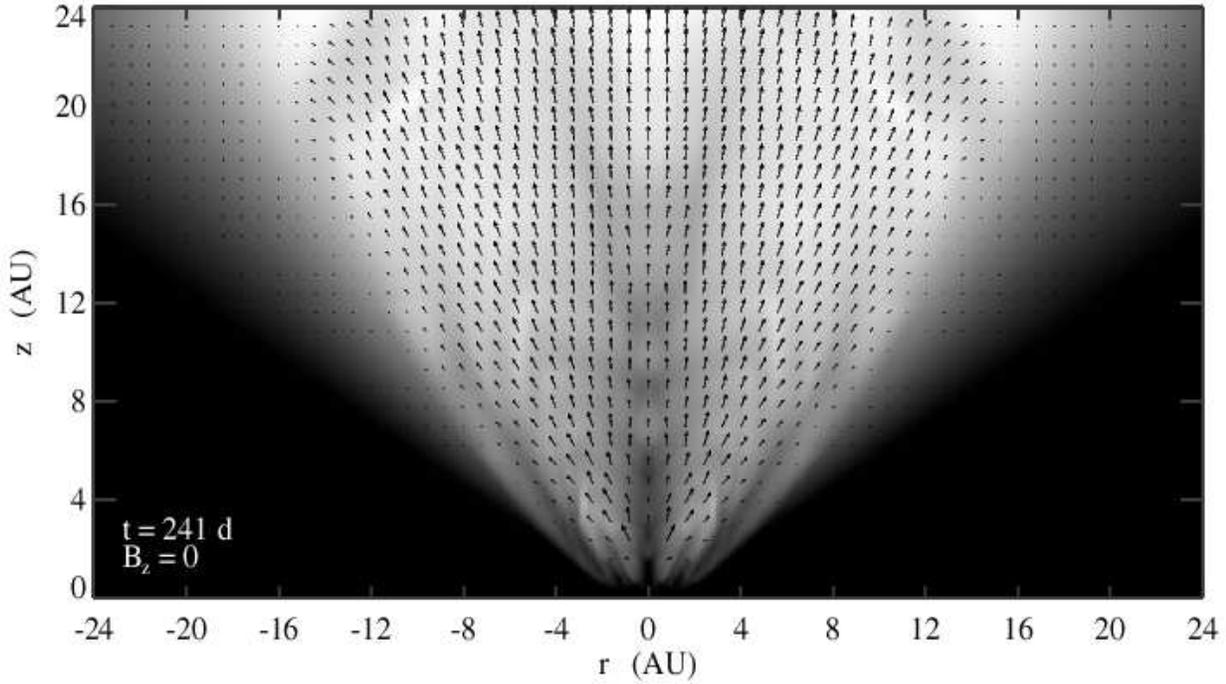}}

\caption[Jet and Wide Angle Flow]{Greyscale image of density 
(logarithmic) and velocity vectors from the case with no vertical
field at 241 days.  Log $\rho > -15.3$ gm cm$^{-3}$ is black, log
$\rho < -19.3$ gm cm$^{-3}$ is white.
\label{fig_densbase}}

\end{figure*}

In the absence of any disc field, the basic properties of the EMI wind
include 1) a relatively high density, narrow jet (the EMI jet) along
the $z$-axis and 2) a wide-angle wind.  These features are apparent in
Figure \ref{fig_densbase}, a snapshot from the tenth simulation box of
the outflow in the $B_z$ = 0 baseline case at 241 days.  The star is
at $r = z = 0$, and the data has been reflected about the $r$ = 0 axis
to better demonstrate the flow.  The Figure also contains data from
the eighth and ninth simulation boxes, which extend to 6 and 12 AU,
respectively, in $z$.  The outflow is highly structured, containing
knots in the jet and `wispy' structures in the wide-angle wind. These
structures are correlated with oscillations of the EMI mechanism.  The
entire flow originates from a region of $\la 30 R_\odot$, where the
inner edge of the disc is located.  By the time of this snapshot,
material originating in the disc or on the star has traveled to $\sim$
16--18 AU, and the flow at larger $z$ is transient (i.e., due to the
initial expansion into a stagnant environment).

\begin{figure*}
\centerline{\includegraphics{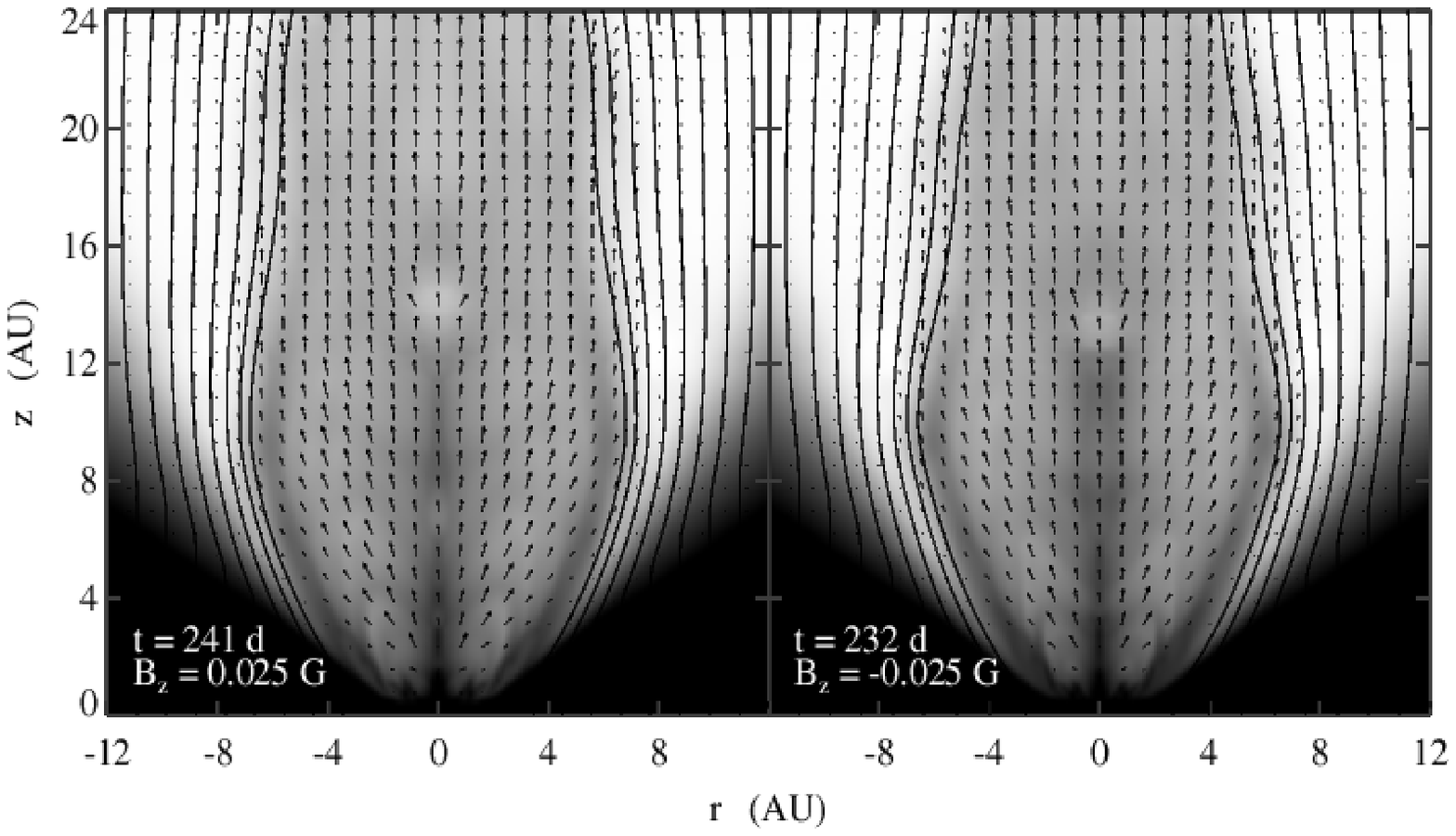}}

\caption[Wind Collimated by 0.025 Gauss Field]{Greyscale images of 
density (logarithmic), magnetic field lines, and velocity vectors from
the case with a 0.025 Gauss vertical field.  The two panels represent
simulations that are identical except for the polarity of the vertical
field (left panel has positive $B_z$, the right has negative, and the
time in days is shown in each panel).  The greyscale is identical to
figure \ref{fig_densbase}, and the line spacing is proportional to
field strength.
\label{fig_densweak}}

\end{figure*}

Figure \ref{fig_densweak} is a snapshot from the $B_z = 0.025$ G (left
panel) and $B_z = -0.025$ G (right panel) cases at 241 and 232 days,
respectively.  It is evident that the wide-angle flow has been
completely collimated by the disc-associated field to a width of 6--7
AU.  In other words, the EMI jet and wide-angle wind have been
combined into a broad, collimated structure (hearafter, the `broad
jet').  Note that the narrow EMI jet is still apparent along the
axis, and a relatively dense shell has formed on the outer edge (in
$r$) of the broad jet, where material traveling at large polar angles
piles up against the disc field before changing direction.  However,
the central EMI jet and the outer shell exists only to a height that
is comparable to the collimation height of the broad jet (at $z
\approx 12$ AU).  A comparison between the left and right panels of
Figure \ref{fig_densweak} reveals that there is almost no difference
in collimation radius for the cases with opposite disc field polarity.
As in figure \ref{fig_densbase}, the main flow has reached to $z \sim$
16--18 AU.  Since this distance is comparable to the width of broad
jet, the final collimation width may become a little larger than
achieved by the time of this snapshot.

\begin{figure*}
\centerline{\includegraphics{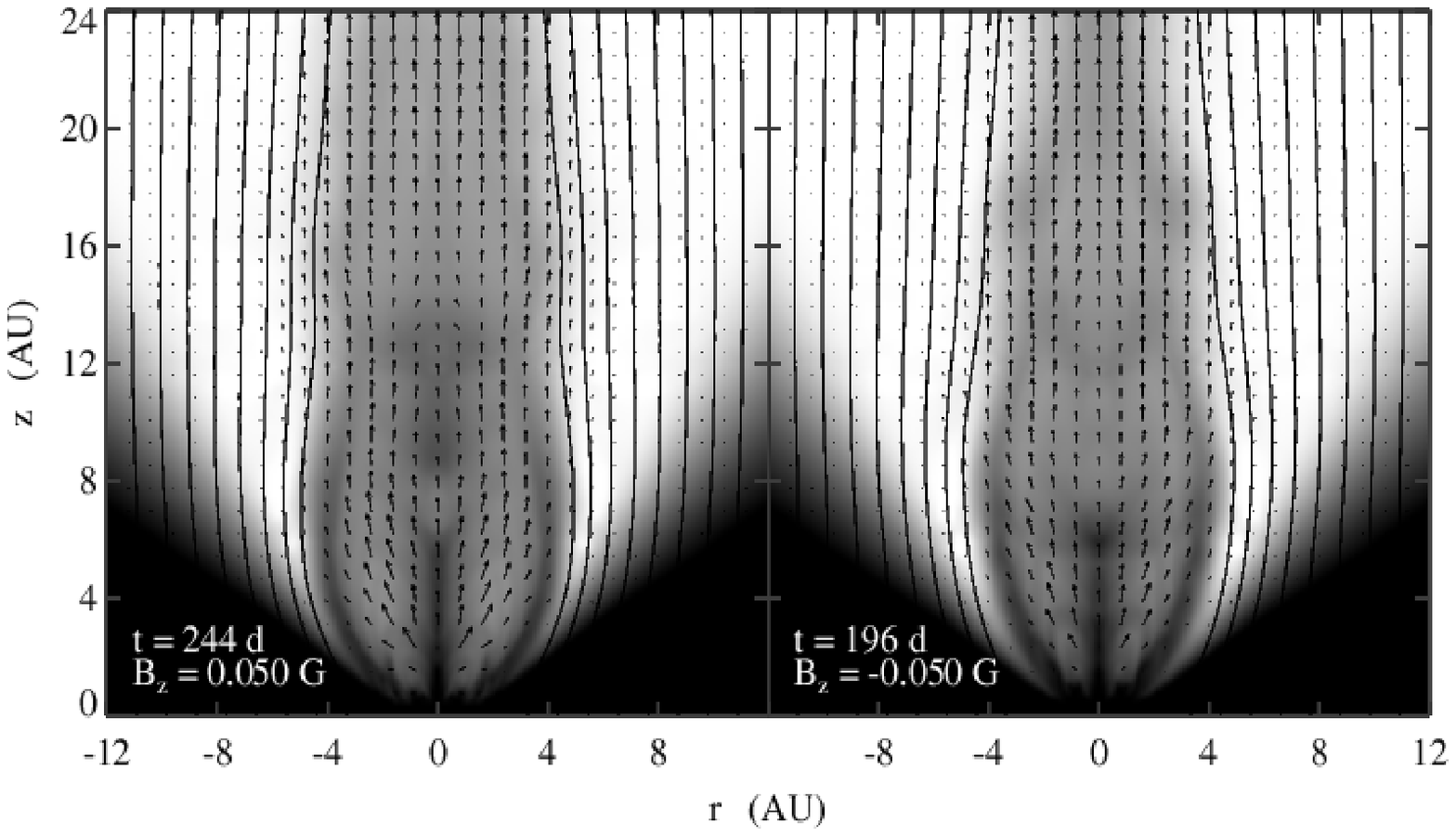}}

\caption[Wind Collimated by 0.05 Gauss Field]{Greyscale images of 
density (logarithmic), magnetic field lines, and velocity vectors from
the case with a 0.05 Gauss vertical field.  The two panels represent
simulations that are identical except for the polarity of the vertical
field (left panel has positive $B_z$, the right has negative, and the
time in days is shown in each panel).  The greyscale is identical to
figure \ref{fig_densbase}, and the line spacing is proportional to
field strength.
\label{fig_densmedium}}

\end{figure*}

Figure \ref{fig_densmedium} is a snapshot from the $B_z = 0.05$ G
(left panel) and $B_z = -0.05$ G (right panel) cases at 244 and 196
days, respectively.  This figure shows the collimation of the entire
central flow to a width of 3--5 AU.  Again, the two panels show that
there is almost no difference in collimation radius between the two
cases of opposite disc field polarity.  As in figure
\ref{fig_densweak}, both panels also contain the central EMI jet and a
dense shell enclosing the broad jet to a height of $z \approx 8$ AU,
where the broad jet becomes collimated.

\begin{figure}
\centerline{\includegraphics[width=22pc]{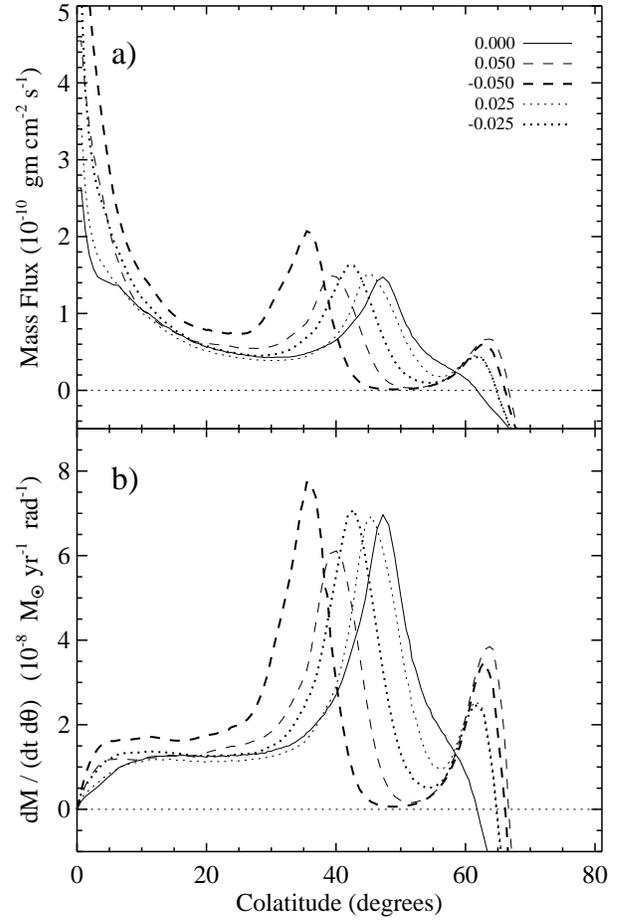}}

\caption[Mass Flux and Outflow Rate]{a) Time-averaged mass flux and b)
time-averaged mass outflow rate per radian ($d\dot M/d\theta$, see
text) versus polar angle, at a constant radius of 5.4 AU.  The solid
line in each panel represents the case without a vertical field ($B_z
= 0$), the dotted lines are for cases with $|B_z| = 0.025$ G, and the
dashed lines are for cases with $|B_z| = 0.05$ G.  Bold lines are for
$B_z < 0$, and the lighter lines are for $B_z > 0$.

\label{fig_massflow}}

\end{figure}

\begin{figure}
\centerline{\includegraphics[width=22pc]{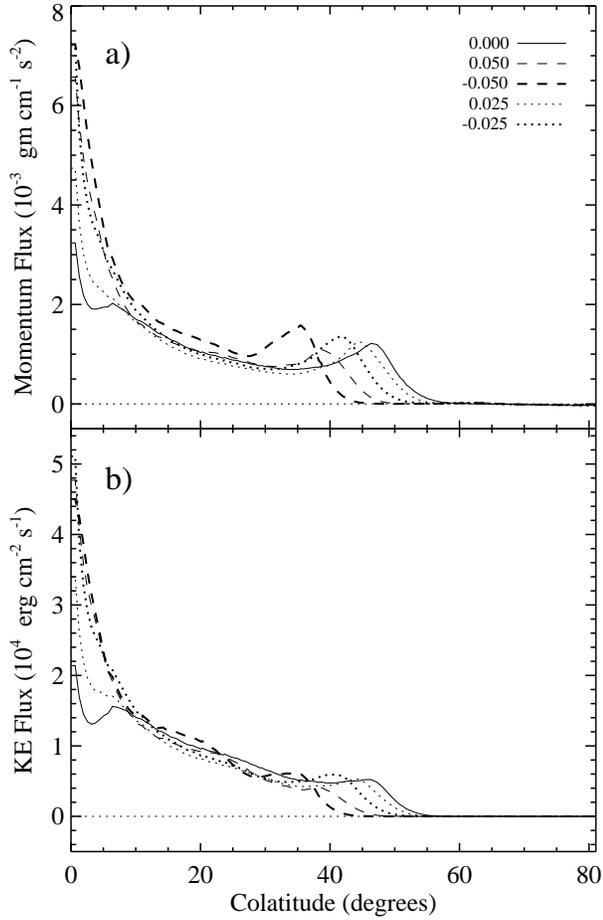}}

\caption[Momentum and Kinetic Energy Flux]{a) Time-averaged linear 
momentum flux and b) time-averaged kinetic energy flux versus polar
angle, at a constant radius of 5.4 AU.  The line styles represent the
same cases as in figure \ref{fig_massflow}.
\label{fig_pkflux}}

\end{figure}

\begin{figure}
\centerline{\includegraphics[width=22pc]{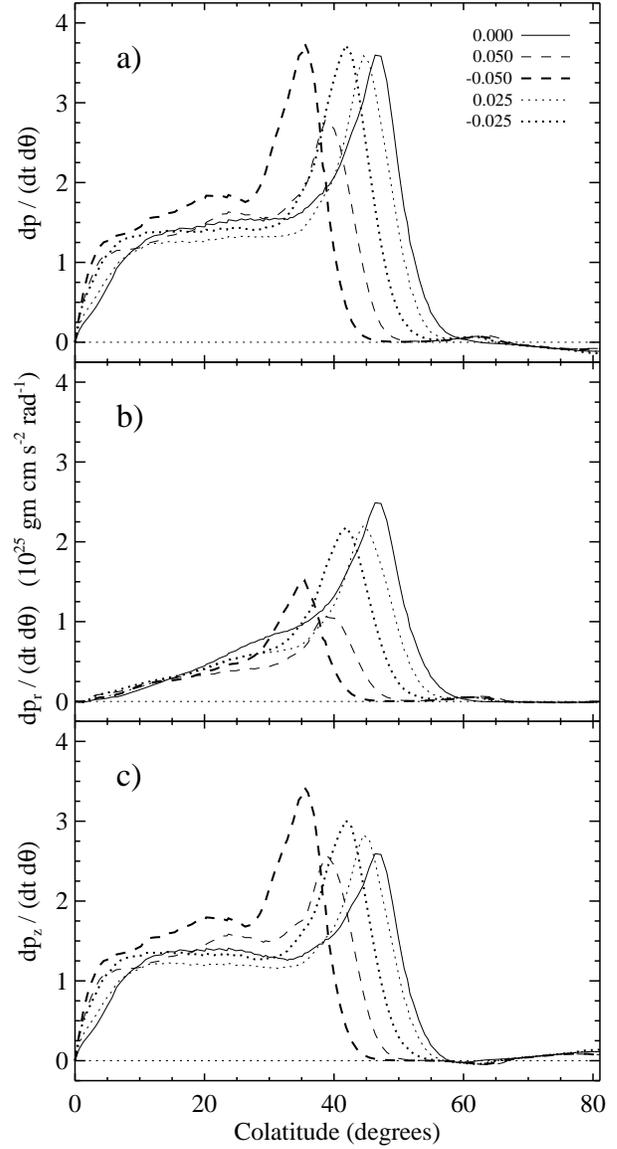}}

\caption[Momentum Outflow Rate]{Time-averaged momentum outflow rate 
(thrust) per radian ($d\dot p/d\theta$, see text) for a) the total,
poloidal momentum, b) the radial component, and c) the vertical
component versus polar angle, at a constant radius of 5.4 AU.  All
three panels are plotted on the same scale.  The line styles represent
the same cases as in figure \ref{fig_massflow}.
\label{fig_pflow}}

\end{figure}

The data plotted in figures \ref{fig_densweak} and
\ref{fig_densmedium} are consistent with the prediction of equation
\ref{eqn_vertjet_sigma} that the collimation radius is inversely
proportional to the disc field strength.  For a more quantitative
analysis, we plot various outflow properties as a function of polar
angle ($\theta$) at a constant spherical radius in figures
\ref{fig_massflow}--\ref{fig_pflow}.  Because the wind is episodic and
highly structured, and because our time sampling for data outputs is
fairly course (2.8 days), it is necessary to represent the outflow
properties as a time-averaged quantities that covers several periods
of the EMI oscillation (which is around 20--30 days).  We therefore
calculate the outflow quantities for each case at 90 gridpoints out in
our eighth simulation box (= 5.4 AU) and average them from 112 days
until the end of the simulation.

Figure \ref{fig_massflow}a shows the mass flux, $\Phi_M(\theta)$, for
all cases.  The existence of the narrow EMI jet at $\theta \la 10$
degrees is represented by the strongly peaked mass flux there for all
cases.  Note that, for $\theta \ga 65^\circ$, $\Phi_M$ becomes
negative, as this represents material inside the accretion disc.
Figure \ref{fig_massflow}b shows the mass outflow rate per
differential polar angle; $d\dot M/d\theta = 2 \pi R^2 \Phi_M(\theta)
\sin\theta$, where $R$ is the spherical radius.  It is evident that,
while $\Phi_M$ is greatest near the pole, most of the mass in the flow
exists at large opening angles ($\theta > 25^\circ$).  For every case,
there is a peak in $\Phi_M$ and $d\dot M/d\theta$ where $\theta$ is
between $25^\circ$ and $55^\circ$.  For larger $B_z$, this peak is
deflected to smaller angles.  The wide-angle flow is not completely
collimated at $z$ = 5.4 AU for any of the cases (see figs.\
\ref{fig_densweak} and \ref{fig_densmedium}), but the deflection of
this peak indicates the onset of collimation.  The cases with positive
$B_z$ are shifted a few degrees less than their negative $B_z$
counterparts.  A positive and negative $B_z$ corresponds to the
stellar magnetosphere having an initially open and initially closed
topology, respectively \citep[see][]{mattea02}.  This suggests that
the open topology is slightly less efficient at collimating at 5.4 AU
(presumably due to the release of magnetic energy from magnetic
reconnection between disc field lines and those carried in the
outflow), though the final collimation radius is not much affected by
the topology of the field (as evident in figs.\ \ref{fig_densweak} and
\ref{fig_densmedium}).  Finally, there is a bump in $\Phi_M$ between
$55^\circ$ and $65^\circ$ for the cases with $|B_z| > 0$.  This
feature is much less prominent in the momentum and kinetic energy (see
figs.\ \ref{fig_pkflux} and \ref{fig_pflow}) of the outflow,
indicating that it has relatively high density and low velocity
compared to the rest of the outflow.  It represents material being
pushed aside by the compression of disc field lines.

The total mass outflow rate can be obtained by integrating $d\dot
M/d\theta$ over a desired range of $\theta$.  For all cases, the
positive mass outflow rate (excluding the bump at $55^\circ < \theta <
65^\circ$ for the $|B_z| > 0$ cases) is $\sim 1.5$--2.0$ \times
10^{-8} M_\odot$ yr$^{-1}$ in one hemisphere.  This agrees very well
with the typical value of $\dot M \sim 3 \times 10^{-8} M_\odot$
yr$^{-1}$ derived from observations of jets \citep{reipurthbally01}.
Note, however, that for the $B_z = 0$ case, the mass loss happens
toward large opening angles, while for the $|B_z| > 0$ cases, the
whole of $\dot M$ is contained in the broad jet.

All cases also show a very strongly peaked EMI jet in the momentum
flux (Fig.\ \ref{fig_pkflux}a), $\rho (v_r^2+v_z^2)^{1/2}
(\mbox{\boldmath$v$} \cdot \hat R)$, and kinetic energy flux (Fig.\
\ref{fig_pkflux}b), $0.5\rho(v_r^2+v_z^2) (\mbox{\boldmath$v$} \cdot
\hat R)$.  The total kinetic energy outflow rate (mechanical
luminosity) of the flow is $\sim 10^{32}$ ergs s$^{-1}$, lower than
the typical value of $\sim 1 L_\odot$ reported by
\citet{reipurthbally01}.  The wide-angle peak in both fluxes of figure
\ref{fig_pkflux} follow the same case-by-case behavior as in figure
\ref{fig_massflow}.  That is, the peak shifts to smaller angles for
stronger $|B_z|$.


Figure \ref{fig_pflow} shows the linear momentum outflow rate (thrust)
per differential polar angle; $d\dot p_x/d\theta = 2 \pi R^2
\sin\theta ~ \times$ $p_x$-flux($\theta$), where $p_x$ is
$m(v_r^2+v_z^2)^{1/2}$ for \ref{fig_pflow}a, $mv_r$ for
\ref{fig_pflow}b, and $mv_z$ for \ref{fig_pflow}c.  The total vertical
momentum outflow rate (\ref{fig_pflow}c) is $\sim 10^{25}$ dynes for
all cases, lower than the typical value of $\sim 10^{26}$ gm cm
s$^{-2}$, reported by \citet{reipurthbally01}.  In all cases, the
total thrust (\ref{fig_pflow}a) is dominated by the vertical thrust
(\ref{fig_pflow}c) for $\theta \la 40^\circ$.  For the baseline case,
the $r$-directed (\ref{fig_pflow}b) and vertical thrust
(\ref{fig_pflow}c) are comparable for $\theta \ga 40^\circ$,
indicating that the flow is moving in a direction of $\sim 45^\circ$
there.  This is because the wide-angle component of the $B_z = 0$ case
is not collimated.  However, \ref{fig_pflow}b demonstrates that, for
increasing $|B_z|$, the total $r$-directed thrust decreases.  This is
because the expanding wind experiences a negative-$r$-directed
{\boldmath $J$} $\times$ {\boldmath $B$} force from the perturbed disc
field.  The peak in vertical thrust (c) shifts to smaller polar angles
for larger $|B_z|$ (indicating the onset of collimation), but the
total vertical thrust (integrated over $\theta$) is roughly the same
for all cases.

The calculation of the outflow thrust (Fig.\ \ref{fig_pflow}) allows
us to apply the results of section \ref{sec_vertjet_isotropic}
quantitatively, though roughly, to the EMI outflow.  Since it is only
the $r$-directed component of the wind that deflects the disc field
before the collimation radius is reached, we will substitute $\dot
M_{\rm w} v_{\rm w}$ of equation \ref{eqn_vertjet_sigma} with $\dot
p_r$ and set $\sigma$ equal to unity.  The collimation radius can then
be estimated as
\begin{equation}
r_{\rm col} \approx {{\sqrt{\dot p_r}} \over {B_z}}.
\label{eqn_vertjet_rcol}
\end{equation}
The $\dot p_r$ (integrated over $\theta$) for the $B_z = 0$ case is
$\sim 6.5 \times 10^{24}$ dynes in one hemisphere.  This predicts a
$r_{\rm col}$ of 4.8 and 9.6 AU for the $|B_z| = 0.05$ and 0.025 G
cases, respectively.  This prediction is slightly larger than the
simulated value for the $|B_z| = 0.025$ cases (see figs.\
\ref{fig_densweak} and table \ref{tab_vertjet_sim2}), but remember
that the simulated flow in that case has not quite reached its full
collimation radius.  Also, for a magnetized central wind (as in the
EMI outflow), there will be an increase in the collimation (decrease
in $r_{\rm col}$) as B$_\phi$ carried in the wind is compressed in the
outer shell \citep[see figs.\ \ref{fig_densweak} and
\ref{fig_densmedium};][]{gardiner3ea02}.


\section{Summary and Discussion} \label{sec_vertjet_discussion}

The relatively high velocity of observed YSO jets requires that jet
material is launched from deep within the gravitational potential well
(i.e., within several stellar radii).  In order to explain the
observed jet widths and their degree of collimation, the jet material
must be launched initially with wide opening angle and become almost
completely collimated at a distance from the star that is comparable
to the width of the jet.  Self-collimation of the wide-angle wind (due
to magnetic fields carried in the wind itself) cannot occur for 100\%
of the wind material; there is always some material leaving at very
wide angles.  This argues for the existence of a collimator that is
external to the central wind.  We have explored one such possibility.

Specifically, we used time-dependent MHD simulations to extend the
work of \citet{kwantademaru88} and demonstrate that any outflow (even
if initially isotropic) will be eventually collimated by a constant,
parallel magnetic field of any strength.  The collimation of the
central wind is inevitable, assuming that the wind is sufficiently
ionized to interact with the field.  This is because the ram pressure
in the wind always decreases with radius for a diverging wind, and the
magnetic pressure is constant.  The collimation radius is roughly
given by the location where the two energy densities balance.

In section \ref{sec_vertjet_application}, we applied this concept to
realistic YSO outflows by first adopting the episodic magnetospheric
inflation (EMI) model \citep{hayashi3ea96, goodson3ea97, goodson3ea99,
millerstone97} for launching the outflows.  We then assumed that the
entire simulation region was initially threaded by a vertical magnetic
field.  The result is that the wide-angle component of the EMI wind is
collimated by disc fields of $\ll 0.1$ Gauss.  In this way, the entire
mass, momentum, and energy budget of the central outflow is channeled
into a relatively fast (100-200 km s$^{-1}$), broad jet with a width
of several AU.  The channeling occurs within a vertical height of the
order of the jet width, consistent with observational constraints (see
\S \ref{sec_vertjet_intro}).

The morphology of the resulting broad jet includes a relatively narrow
density enhancement along the axis.  This feature is the narrow jet
produced by the EMI mechanism alone (i.e., even when there is no disc
field).  The collimation of the wide-angle wind results in a
relatively high density shell of shocked, outflowing material
surrounding the flow out to at least the height of collimation.  The
EMI jet and shocked shell together resemble a `pitchfork' in
projection (see figs.\ \ref{fig_densweak} and \ref{fig_densmedium}).

\citet{bally3ea03} recently suggested that x-ray emission from a
region 0\farcs5 to 1\farcs0 west-southwest of the YSO L1551 IRS 5 may
be explained by shocks associated with the collimation of the main
outflow from that region.  Using kinematic information from IR slit
observations of that outflow, \citet{pyoea02} demonstrated the
existence of two different flows (not including the molecular outflow
discovered by \citealp{snell3ea80}): an ionized, highly collimated jet
moving at $\sim 440$ km s$^{-1}$, and a partially ionized, wide-angle
wind moving at $\sim 200$ km s$^{-1}$.  The wide-angle wind shows a
kinematic signature of increased collimation as it moves further from
the source.  \citet{pyoea02} also noted a pitchfork-like morphology in
an I band image of the outflow, which they suggested could represent
the limb-brightened edges of the wide-angle wind under
collimation. Note, however, that the southernmost of the pitchfork
`tines' has been interpreted as a separate jet from a companion star
\citep{fridlundliseau98, hartiganea200}.



In this paper, we have assumed the special geometry of an
axis-aligned, vertical field, which requires that the field is
associated with the accretion disc.  This field could represent a
field carried in a disc wind (moving more slowly than the central
wind), though we don't follow such a wind in our simulations.
Specifically, if the disc is magnetized, and if there is a magnetized
outflow from the disc, the field will roughly follow the wind
streamlines in the poloidal plane.  Indeed, the predicted poloidal
field geometry for disc wind launching models
\citep[e.g.,][]{uchidashibata85, shibatauchida86, shibatauchida90,
ouyedpudritz97, ustyugovaea99} is essentially vertical near the
rotation axis.

Disc wind models also predict that the field has a helical structure,
due to the winding up of field lines by the rotation of the disc.  The
azimuthal field, added in this way, increases the collimation of the
disc wind \citep{blandfordpayne82, sakurai85}, and would therefore
further constrain any central, wide-angle flow that interacts with it.
Due to the long time-scale for the winding of disc field compared to
the time-scale for the launching and propagation of the central wind by
the EMI mechanism, our simulations do not capture the effect of the
azimuthal component expected in the disc-associated field.

Also, due to differential rotation and possibly turbulence in a real
accretion disc, it is reasonable to assume that any field present in
the disc may be disordered, and possibly chaotic
\citep[e.g.,][]{blandfordpayne82, hawleybalbus92}.  A disc wind would
therefore be threaded by a magnetic field with direction reversals at
irregular intervals (i.e., flux tubes with opposite polarity).  Such
direction reversals do not affect the qualitative behavior of the
field \citep[see, e.g.,][]{tsinganosbogovalov00}.  That is, whether
the vertical field has a constant or chaotic polarity, it will always
act to collimate the flow (see also \citealp{li02} for similar
description with chaotic field).

The interaction of a wide-angle wind with an external, disc-associated
magnetic field (as presented here) is not the only possible
large-scale collimation mechanism for YSO outflows.  A few authors
\citep{franknoriegacrespo94, frankmellema96, lishu96, mellemafrank97,
delamarter3ea00, gardiner3ea02} have discussed the interaction of
wide-angle flows with an unmagnetized environment \citep[see
also][]{leryea02}.  They have found that the wide winds may be
completely collimated by shock focusing, especially when azimuthal
magnetic fields are present in the wide-angle wind.  It is not clear
what the observational signature of such an interaction would be
\citep[for possible detections, see][]{mattbohm02, bally3ea03}, but it
takes place relatively close to the source and therefore may be
difficult to disentangle from the complex emitting environment there.

\section*{Acknowledgments}

This research was supported by NSF grant AST-9729096 and by the
National Science and Engineering Research Council of Canada (NSERC),
McMaster University, and the Canadian Institute for Theoretical
Astrophysics through a CITA National Fellowship.





\begin{table}

 \caption{Isotropic wind simulation parameters. \label{tab_vertjet_sim1}}

 \begin{tabular}{@{}lcccc}

  \hline

Case Notes &
$\sigma_*$ & 
${\rho_*} \over {\rho_{\rm a}}$ & 
${v_{\rm w}} \over {c_{\rm s,w}}$ & 
${v_{\rm A,a}} \over {c_{\rm s,a}}$ \\

  \hline

baseline 			& 0.0087 & 10 & 5 & 5 \\
weak $B_z$ 			& 0.0043 & 10 & 5 & 5 \\
very weak $B_z$ 		& 0.0022 & 10 & 5 & 5 \\
strong $B_z$ 			& 0.0174 & 10 & 5 & 5 \\
low $\rho_{\rm a}$		& 0.0087 & 100 & 5 & 5 \\
high $\rho_{\rm a}$		& 0.0087 & 1 & 5 & 5 \\
low $\rho_{\rm a}$, weak $B_z$ 	& 0.0043 & 100 & 5 & 5 \\
cold wind 			& 0.0087 & 10 & 10 & 5 \\
cold ambient 			& 0.0087 & 10 & 5 & 10 \\
$B_z$ = 0			& 0.0000 & 10 & 5 & 5 \\


  \hline

 \end{tabular}

\end{table}

\begin{table}

 \caption{YSO simulation parameters. \label{tab_vertjet_sim2}}

 \begin{tabular}{@{}rcl}

  \hline

$B_z$ &
$t_{\rm f}$ & 
Notes: \\ 

(Gauss) & 
(days) & \\

  \hline

0.000  & 241 & no disc field \\
0.025  & 241 &  $r_{\rm col} \ga $ 6--7 AU\\
$-0.025$ & 232 &  $r_{\rm col} \ga $ 6--7 AU\\
0.050  & 244 &  $r_{\rm col} \approx $ 3--5 AU\\
$-0.050$ & 196 &  $r_{\rm col} \approx $ 3--5 AU\\

  \hline

 \end{tabular}

\end{table}


\label{lastpage}

\end{document}